\documentstyle[psfig,prb,aps,twocolumn]{revtex}
\begin{document}
\draft

\title{Matrix product states approach to the Heisenberg
ferrimagnetic spin chains}

\author{A. K. Kolezhuk,$^{1,2}$ H.-J. Mikeska,$^1$
and Shoji Yamamoto$^{1,3}$}
\address{$^1$Institut f\"{u}r Theoretische Physik,
Universit\"{a}t Hannover, 30167 Hannover, Germany}
\address{\cite{perm}$^2$Institute of Magnetism, National Academy of
Sciences of Ukraine, 252142 Kiev, Ukraine }
\address{\cite{perm}$^3$Department of Physics, Faculty of Science, Okayama
University, Okayama 700, Japan}
\date{\today}

\maketitle
\widetext
\begin{abstract}
\hskip 0.75 in\parbox[t]{5.5in}{%
We present a new method of constructing the matrix product (MP)
states and apply it to the ground state of the Heisenberg spin
chain with alternating spins $1$ and $1\over2$ and antiferromagnetic
exchange interactions between nearest neighbors (the simplest
example of a quantum ferrimagnet). The elementary matrix state is
constructed in a way which ensures given transformational
properties under rotations, which allows one to fix the total
spin and its $z$-projection of the entire MP wavefunction.
We compare the variational MP results with the numerical results
obtained through a Quantum Monte Carlo method; the agreement is
found to be within 0.4\% for the ground state energy and 5\% for
the correlation functions.
}
\end{abstract}

\narrowtext


Recently, there has been a considerable progress in the study of
one-dimensional (1d) quantum spin chains by means of the
so-called {\em matrix product} (MP) states technique.
\cite{Fannes+89,Klumper+91,Fannes+92,Klumper+92-93} For the first
time, they were introduced \cite{Fannes+89} as a convenient
representation
of the valence bond  states\cite{AKLT} (VBS) for integer-spin chains.
In terms of the original
spin states, VBS wavefunctions are highly non-local, but they
factorize in terms of matrix states which considerably simplifies
all calculations.  Shortly after that, MP states have proved to
be a useful tool for constructing new classes of spin
Hamiltonians with exactly known ground states,
\cite{Klumper+91,Fannes+92,Klumper+92-93,NiggZitt96}  for
variational study of the ground state properties of spin chains
\cite{SchadZitt95} and spin ladders, \cite{Brehmer+96} and also
for the study of elementary excitations which are solitons in the
string order. \cite{TotsukaSuzuki95,NeuMik96}

In this paper, we present a new method of constructing the matrix
product (MP) states with fixed quantum numbers, namely the total
spin and its $z$-projection, and demonstrate its use on the
example of the Heisenberg spin chain with antiferromagnetic
nearest neighbor interaction and alternating spins $1$ and
$1\over2$. This system, according to the Lieb-Mattis theorem,
\cite{LiebMattis62} has a degenerate multiplet of ground states
with the total spin $S_{\text{tot}}=L/2$; here $L$ is the number
of unit cells, each cell consisting of spins 1 and
$1\over2$. In each of the ground states the rotational symmetry
is spontaneously broken, and the long-range magnetic order
exists;\cite{BreMikYam96} thus, one may consider  this model as
the simplest example of a quantum ferrimagnet. It should be
mentioned that there exist real magnetic materials belonging to
the family of Cu(II)Ni(II) complexes, which are well described by
the above model, see Ref.\ \onlinecite{Kahn87} and references
therein.
For this model, we construct a MP wavefunction which has only 
two variational parameters, and show that our variational
results for
the ground state properties are in a very good agreement with the
numerical data obtained by means of 
%
\newpage\vbox to 1.9in{\noindent}\noindent
%
the Quantum Monte Carlo (QMC) technique.


{\em Matrix states with given quantum numbers.\/}
First, let us recall some basic facts about the MP states. A
matrix product state is defined as
\begin{equation}
|\Omega\rangle =\mbox{Tr}\,(g_1 g_2 \cdots g_L) \,,
\label{MP}
\end{equation}
where the elementary matrix states $g_i$ are matrices composed
from the spin states of the
$i$-th magnetic elementary cell. The dimension
of the elementary matrix depends on the problem: for example,
translationally invariant VBS state for spin-$S$ chain can be
represented with the help of $(S+1)\times(S+1)$ matrices.
\cite{TotsukaSuzuki95} In the following, we will assume
that $g_i$ is a square matrix.
It is worthwhile to note that since the definition of (\ref{MP})
involves the trace sign, the MP state does not depend on the
representation of the elementary matrices, i.e., any {\em global}
unitary transformation $g_i \mapsto U g_i U^\dagger$ keeps
$|\Omega\rangle$ invariant.

Let us demand that the elementary matrix $g$ has certain
transformational properties. Namely, we would like to construct a
multiplet of matrices $g^{jm}$ transforming according to the
$\bar{{\cal D}}^j$ representation of the rotation group under
rotations $\widehat T_R$ :
\begin{equation}
\widehat T_R g^{jm}\simeq \sum_{m'} {\cal D}^j_{m'm}(R)\, g^{jm'}
\,,
\label{trg}
\end{equation}
where the sign $\simeq$ means unitary equivalence.
Further, assume that the desired dimension of the elementary
matrix is ($N$+$1$)$\times$($N$+$1$) and that we know the complete set of
wavefunctions of the elementary cell
$\{|\psi_{\lambda\mu}\rangle\}$, $\lambda$ being the total spin
and $\mu$ its $z$-projection. One can choose a basis in the
matrix space consisting of the matrices
$X^{kq}=(T^{kq})^\dagger$, $k=0\ldots N$, $q=-k\ldots k$.
Here $T^{kq}$ are the
($N$+$1$)$\times$($N$+$1$) matrix representations of irreducible tensor
operators\cite{Yutsis+62,Blum81}
transforming according to the ${\cal D}^k$
representation of the rotation group; defining
$X^{kq}=(T^{kq})^\dagger$ just ensures that $X^{kq}$ transform
according to the conjugate representation $\bar{{\cal D}}^k$,
like usual wavefunctions.
In the simplest case $N$=$1$ (2$\times$2 matrix space) the
$T^{kq}$
matrices are just the identity matrix $1\!\!1$ and the
Pauli matrices
$\sigma^0=\sigma_z$, $\sigma^{\pm}= \mp 1/\sqrt{2}\,
(\sigma_x\pm i\sigma_y)$:
\begin{equation}
T^{00}=1\!\!1\,,\quad T^{10}=\sigma^0,
\quad T^{1,\pm1}=\sigma^{\pm};
\label{T2by2}
\end{equation}
explicit expressions for $T^{kq}$ in 3$\times$3 space are given
below.
Then, our basis transforms as
\begin{equation}
\widehat T_R X^{kq}=\sum_{q'} {\cal D}^k_{q'q}(R)\, X^{kq'} \,,
\label{trx}
\end{equation}
which on the other hand amounts to a unitary transformation
$\widehat T_R X^{kq} = U(R)\, X^{kq} U^\dagger(R)$ and does not have
any effect on the MP wavefunction.

It is easy to show that the following construction satisfies our
requirement (\ref{trg}):
\begin{equation}
g^{jm}=\sum_{k\lambda} c^{k\lambda}_j \sum_ {q\mu}
\langle jm | kq,\lambda\mu\rangle
X^{kq} |\psi_{\lambda\mu}\rangle\,.
\label{gen}
\end{equation}
Here $\langle jm | kq,\lambda\mu\rangle$ are the Clebsch-Gordan
coefficients, and $c^{k\lambda}_j$ are arbitrary constants. In
fact, we used the usual quantum mechanical addition rules for
angular momenta, demanding that $g$ has certain quantum numbers
$j,m$ characterizing the ``hyperspin" defined in an extended
space which is a direct product of ($N$+$1$)$\times$($N$+$1$)
matrix space and the Hilbert space of the cell.

Then, we can construct a translational invariant matrix state
with given hyperspin $J=Lj$ and its $z$-projection $M=J$:
\begin{equation}
G^{JJ}
=g^{jj}_1 g^{jj}_2 \cdots g^{jj}_L \,.
\label{MPJM}
\end{equation}
The matrix state $G^{JJ}$ belongs to the multiplet transforming
according to $\bar {\cal D}^J$ and thus has the same structure
(\ref{gen}) as $g^{JJ}$, but now $|\psi_{\lambda\mu}\rangle$ have
the meaning of many-body wavefunctions  of an open
spin chain (their total spin $\lambda=|J-N|,\ldots, J+N$ can
differ from $J$ because of the edge effects). Taking the trace
$|\Omega\rangle=\mbox{Tr}\,(G^{JJ})$, we pick up only the
$X^{00}$ component, i.e. the state $|\psi_{JJ}\rangle$
describing a chain with periodic boundary conditions. Other
states of the multiplet with $M<J$ can be obtained from
$|\psi_{JJ}\rangle$ by applying $\widehat{S}^-_{\text{tot}}$.

It should be mentioned that the construction (\ref{gen}) allows
several different values of $j$ even for the same size of the
magnetic elementary cell; the number of possibilities grows fast
with increasing the cell size, when more and more complicated magnetic
structures with longer period are allowed. One can also construct
various translationally non-invariant `excited' MP states, but
this goes beyond the scope of the present paper.

One can see that known MP states for
VBS models \cite{TotsukaSuzuki95} and spin-$1\over2$
ladders \cite{Brehmer+96} indeed follow the
general structure described above. For example, the elementary
matrix for the spin-1 AKLT state can be written in the form
\[
g_{\text{AKLT}}= {1/\sqrt{3}}\left\{
(\sigma^{+})^\dagger |-\rangle +
(\sigma^{-})^\dagger |+\rangle -
(\sigma^0)^\dagger |0\rangle \right\}\,,
\]
which is exactly $g^{00}$ in our notation, and four components of
the matrix $G^{00}$ defined by (\ref{MPJM})  describe a singlet and the
Kennedy triplet states of the open spin-1 chain.

In the spin-2 case one possibility is to construct $T^{1m}$ as
the matrix representations of $S$=$1$ spin operators, and $T^{2m}$ can be
expressed\cite{Blum81} as bilinear forms of $T^{1m}$. This
yields the following expressions for $T^{kq}$ in 3$\times$3 space:
\begin{eqnarray}
&& T^{10}={1\over\sqrt 2}
\left(\begin{array}{rrr} 1& 0& 0\\
            0& 0& 0\\
            0& 0& -1\end{array}\right),\;\;
T^{1,+1}={1\over\sqrt 2}
\left(\begin{array}{rrr} 0& -1& 0\\
            0& 0& -1\\
            0& 0& 0\end{array}\right),\nonumber\\
&& T^{20}={1\over\sqrt 6}
\left(\begin{array}{rrr} 1& 0& 0\\
            0& -2& 0\\
            0& 0& 1\end{array}\right), \;\;
 T^{2,+1}={1\over\sqrt 2}
\left(\begin{array}{rrr} 0& -1& 0\\
            0& 0& 1\\
            0& 0& 0\end{array}\right),\nonumber\\
&& T^{2,+2}=
\left(\begin{array}{rrr} 0& 0& 1\\
            0& 0& 0\\
            0& 0& 0\end{array}\right).
\label{T3by3}
\end{eqnarray}
The matrix $T^{00}$ is proportional to the identity matrix $1\!\!
1$, and the relationship $T^{j,-m}=(-1)^m (T^{jm})^\dagger$
holds. If the elementary cell consists of a single $S$=$2$ spin,
then the elementary matrix $g^{00}$ constructed according to
(\ref{gen}) is $g_{S=2}=\sum_m (T^{2m})^\dagger
|\psi_{2m}\rangle$, which coincides with the $S$=$2$ VBS
matrix;\cite{TotsukaSuzuki95} nine components of $G^{00}$
describe degenerate ground states of the open spin-2 VBS chain.

In case of a spin-$1\over2$ ladder
the elementary cell consists of two spins, and according to
(\ref{gen}) one gets the ansatz used in Ref.\ \onlinecite{Brehmer+96}:
\[
g_{\text{ladder}} = c^{00}_0\cdot
1\!\!1 \cdot |s\rangle +c^{11}_0 g_{\text{AKLT}}\,.
\]
Here $|s\rangle $ denotes the singlet state, and the states
$|\mu\rangle$ in $g_{\text{AKLT}}$, $\mu=0,\pm1$ now should be
interpreted as the triplet states of two $S$=$1\over2$ spins.

Our ansatz (\ref{gen}) allows one to construct MP states with
spontaneously broken rotational symmetry (nonzero total spin
$J$), which is practically important for the case of
ferrimagnets. In the next Section we apply our formalism to the
simplest model of a quantum ferrimagnet and argue that it really
provides a good description of its ground state properties.

\begin{figure}
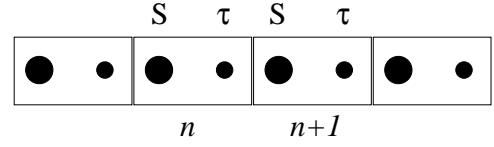

\caption{\label{fig:chain}
The Heisenberg spin chain  with alternating
spins $1$ (large circles) and $1\over2$ (small circles);
boxes show the unit cells.}
\end{figure}

{\em Application to the Heisenberg model with alternating
spins $1$ and ${1\over2}$.\/}
We consider the model described by the Hamiltonian
\begin{equation}
\widehat H= \sum_n (\mbox{\boldmath$S$\unboldmath}_n 
\mbox{\boldmath$\tau$\unboldmath}_n +
\mbox{\boldmath$\tau$\unboldmath}_n \mbox{\boldmath$S$\unboldmath}_{n+1})\,
\label{ham}
\end{equation}
where $\mbox{\boldmath$ S$\unboldmath}_n$ and
$\mbox{\boldmath$\tau$\unboldmath}_n$ are respectively spin-1 and
spin-$1\over2$ operators at the $n$-th elementary magnetic cell
(with $S^z$ eigenstates denoted as $(+,0,-)$ and
$(\uparrow,\downarrow)$, respectively). The elementary cell
consists of spins 1  and $1\over2$, the exchange interaction
exists only between nearest neighbors, and all bonds are
antiferromagnetic and of the same strength which is set to unity
(see Fig.\ \ref{fig:chain}).

According to the Lieb-Mattis theorem,\cite{LiebMattis62}
the ground state of this
model has total spin $S_{\text{tot}}=L/2$, where $L$ is the number of
elementary cells. The rotational symmetry is
spontaneously broken in the ground state, which leads to a
ferrimagnetic long-range order. Thus, we have to construct the
elementary matrix $g$ with $j=1/2$, $m=1/2$. The complete set
of cell wavefunctions $\psi_{\lambda\mu}$ contains one doublet
($\lambda=1/2$) and one quartet ($\lambda=3/2$):
\begin{eqnarray}
&&|\psi_{{1/2},{1/2}}\rangle  = \sqrt{2/3}\, |+\downarrow\rangle
-1/\sqrt{3}\, |0\uparrow\rangle\,,\nonumber\\
&&|\psi_{{1/2},-{1/2}}\rangle  = -\sqrt{2/3}\, |-\uparrow\rangle
+1/\sqrt{3}\, |0\downarrow\rangle\,,\nonumber\\
&&|\psi_{{3/2},{3/2}}\rangle = 
|+\uparrow\rangle\,,\quad
|\psi_{{3/2},-{3/2}}\rangle = |-\downarrow\rangle\,,\\
\label{wfset}
&&|\psi_{{3/2},{1/2}}\rangle = \sqrt{2/3}\, |0\uparrow\rangle
+1/\sqrt{3}\, |+\downarrow\rangle\,,\nonumber\\
&&|\psi_{{3/2},-{1/2}}\rangle  =\sqrt{2/3}\, |0\downarrow\rangle
+1/\sqrt{3}\, |-\uparrow\rangle\,,\nonumber
\end{eqnarray}

Let us choose the simplest case of 2$\times$2
matrix space, then the basis is $X^{00}=1\!\!1$,
$X^{10}=\sigma^0$, $X^{1,\pm1}=-\sigma^{\mp}$ (cf. Eq.\
(\ref{T2by2})), and,
according to (\ref{gen}), the most general form of the elementary
matrix with $j=m=1/2$ is
\begin{eqnarray}
&& g=c_1 M^{0,{1/2}} + c_2 M^{1,{1/2}}
+ c_3 M^{1,{3/2}}\,, \nonumber\\
&& M^{0,1/2}=1\!\!1 \, |\psi_{1/2,1/2}\rangle\,,\\
\label{ansatz}
&& M^{1,1/2}=-1/\sqrt{3}\,\sigma^0\,|\psi_{1/2,1/2}\rangle
-\sqrt{2/3}\,\sigma^{-1}\,|\psi_{1/2,-1/2}\rangle\,,
\nonumber\\
&& M^{1,3/2}=-1/\sqrt{3}\,\sigma^0\,|\psi_{3/2,1/2}\rangle
-1/\sqrt{6}\,\sigma^{-1}\,|\psi_{3/2,-1/2}\rangle\nonumber\\
&&\quad\quad\quad
-1/\sqrt{2}\,\sigma^{+1}\,|\psi_{3/2,3/2}\rangle\,.\nonumber
\end{eqnarray}
Thus, $g$ contains two independent variational parameters
$u=c_1/c_3$ and $v=c_2/c_3$ which are assumed to be the same
throughout the chain. By the construction, the variational MP
state $|\Omega\rangle =\mbox{Tr}\,(g_1 g_2 \cdots g_L)$ has total
spin $S_{\text{tot}}=L/2$, i.e., the same as the true ground state.

\begin{figure}
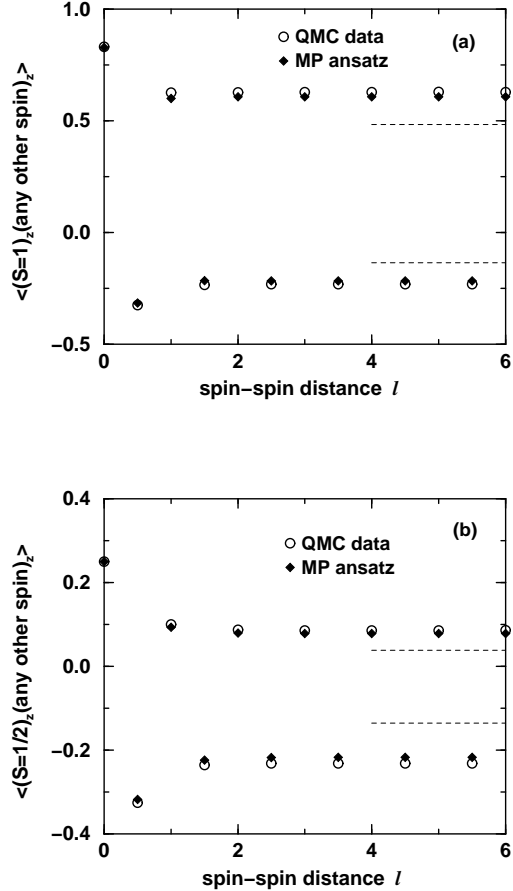

\caption{\label{fig:corr}
Longitudinal spin correlation functions $K^{zz}_{ss'}(l)$, where
$s$ and $s'$ can be either $S$=$1$ or $S$=$1\over2$, and $l$ is
the distance between spins in unit cell lengths, so that for
integer (half-integer) $l$ the correlation is measured between
spins of same (different) type, respectively; (a) correlations
between $S$=$1$ and any other spin; (b) correlations between
$S$=$1\over2$ and any other spin. The dashed line shows the
asymptotic value predicted by the spin wave
theory.\protect\cite{BreMikYam96} The QMC data were obtained at
$L=32$. The numerical uncertainty is less than $0.002$, which is
much smaller than the symbol size.}
\end{figure}

The quantum averages can be calculated using the
transfer matrix technique; for example, for any operator
$\widehat F$ acting on the states of only one unit cell one
has
$\langle \Omega| \widehat F |\Omega\rangle=\mbox{Tr}\,(G^{L-1}
G_F)/\mbox{Tr}\,(G^L)$,
where  $G=g^*\otimes g$ is the transfer matrix, $G_F=g^*\otimes
(\widehat F g)$, and the sign $\otimes$ means an outer
matrix product; further details can be found in
Refs.\ \onlinecite{Klumper+91,TotsukaSuzuki95}. The resulting
expression for the ground state energy per unit cell is
rather lengthy but still manageable,
\begin{eqnarray}
&& E_{\text{var}} =  (1/18 Q)( A + B/Z)\,, \\
\label{varE}
&& Z = [12v^2(u^2+1)+3]^{1/2},\quad
Q = (3u^2+v^2+1+Z)^2,
\nonumber\\
&& A = -180u^4-360u^2v^2-20v^4\nonumber\\
&&\quad\quad -180 \sqrt{2}u^2v +36\sqrt{6}uv^2
-40\sqrt{2}v^3\nonumber\\
&&\quad\quad  -108u^2 +36\sqrt{3}uv -96v^2 +27\sqrt{6}u
-29\sqrt{2}v +4\,,\nonumber\\
&& B = -1440u^4v^2-480u^2v^4 -180 \sqrt{2}u^4v
+108\sqrt{6}u^3v^2\nonumber\\
&&\quad -540 \sqrt{2}u^2v^3 +36\sqrt{6}uv^4
-1404u^2v^2 +288\sqrt{3}uv^3 \nonumber\\
&&\quad -108v^4 -210\sqrt{2}u^2v +234\sqrt{6}uv^2 
-72\sqrt{2}v^3 \nonumber\\
&&\quad -144u^2 +108\sqrt{3}uv -108v^2 +81\sqrt{6}u
-27\sqrt{2}v +18\,.
\nonumber
\end{eqnarray}
Its numerical minimization gives
$E_{\text{var}}^{\text{min}}\equiv E_{\text{MP}}=-1.449$ (at $u=-1.303$,
$v=1.079$). Using this solution, one can 
\vskip 0.1in
\mbox{\hspace{0.3in}\psfig{figure=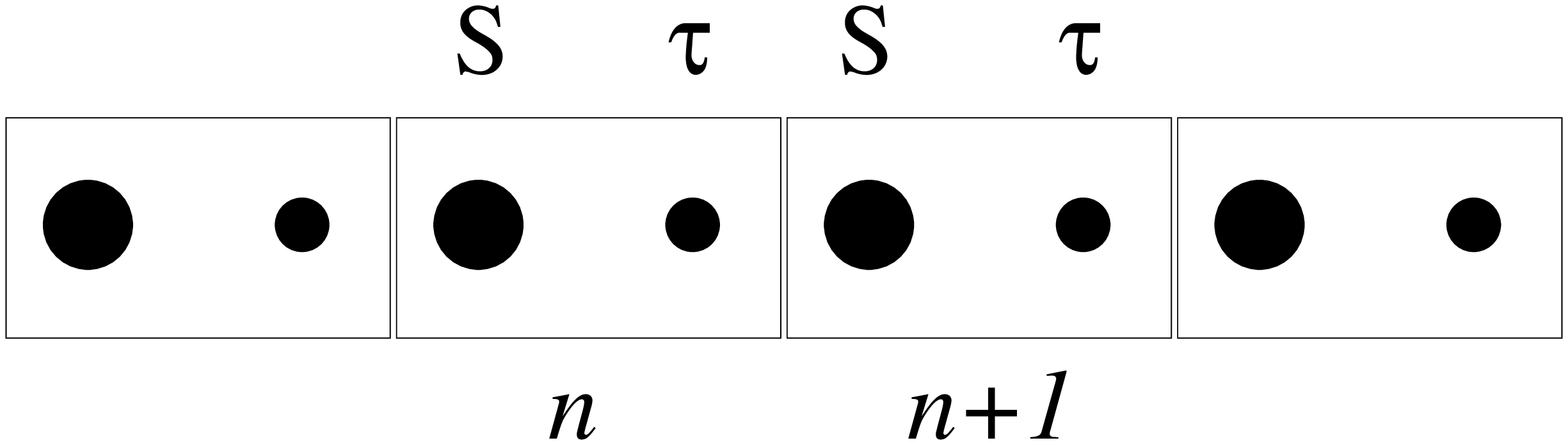,height=0.7in,angle=-90.}}
 \vskip 0.12in\nopagebreak
  \noindent\parbox[t]{3.40in}{\protect\small
FIG.\ \ref{fig:chain}.
The Heisenberg spin chain  with alternating
spins $1$ (large circles) and $1\over2$ (small circles);
boxes show the unit cells.}
\vskip 0in
\mbox{\hspace{-0.2in}\psfig{figure=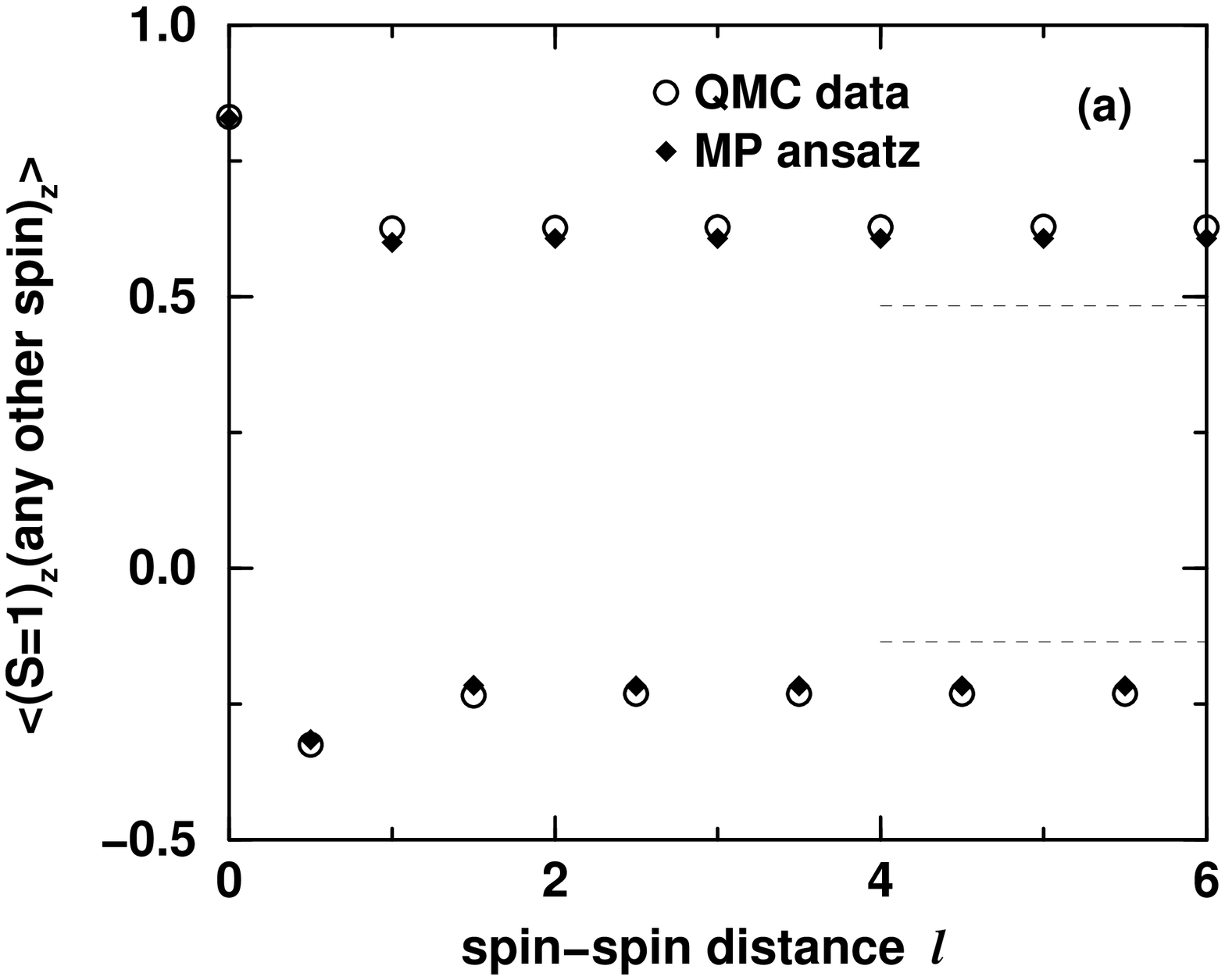,height=2.75in,angle=-90.}}
\vskip -0.2in
\mbox{\hspace{-0.2in}\psfig{figure=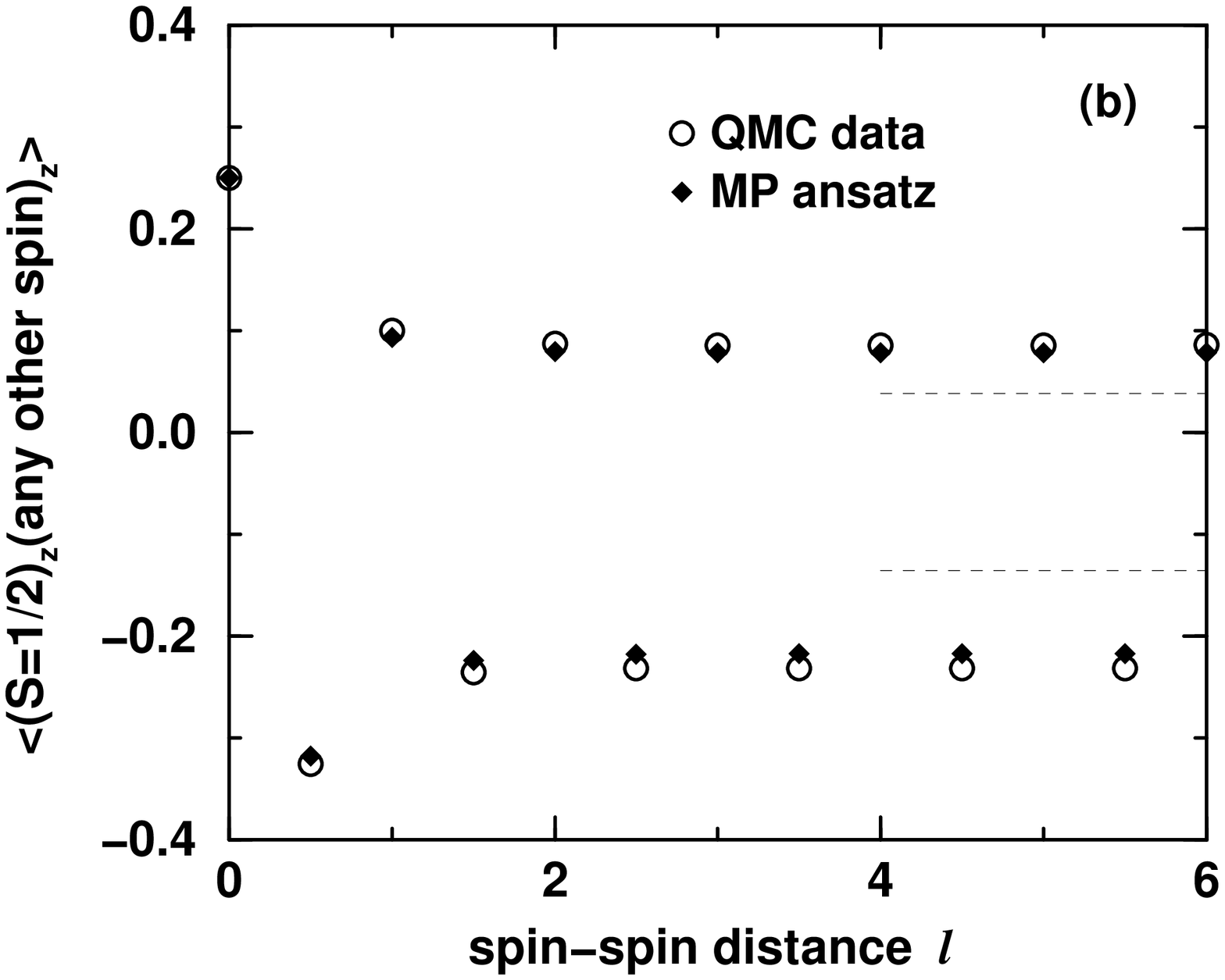,height=2.75in,angle=-90.}}
 \vskip 0.05in\nopagebreak
  \noindent\parbox[t]{3.40in}{\protect\small
FIG.\ \ref{fig:corr}.
Longitudinal spin correlation functions $K^{zz}_{ss'}(l)$, where
$s$ and $s'$ can be either $S$=$1$ or $S$=$1\over2$, and $l$ is
the distance between spins in unit cell lengths, so that for
integer (half-integer) $l$ the correlation is measured between
spins of same (different) type, respectively; (a) correlations
between $S$=$1$ and any other spin; (b) correlations between
$S$=$1\over2$ and any other spin. The dashed line shows the
asymptotic value predicted by the spin wave
theory.\protect\cite{BreMikYam96} The QMC data were obtained at
$L=32$. The numerical uncertainty is less than $0.002$, which is
much smaller than the symbol size.}
\vskip 0.15in
\noindent
also
calculate average spin values $\langle s^z_i\rangle$ and
spin correlation functions $\langle s^z_i
s'^z_{i+n}\rangle$, where $s,s'\in \{ S,\tau \}$. Correlation
functions behave as $a+be^{-n/\xi}$, the correlation length $\xi$ being
extremely short, $\xi=0.365$ in the unit cell lengths. In the regular
spin wave (SW) theory the decay is exponential only
asymptotically, and $\xi\approx0.7$;\cite{BreMikYam96} the spin
reduction is also overestimated in the SW calculation.

The results were compared with the numerical data obtained with
the quantum Monte Carlo (QMC) method based on the Trotter-Suzuki
decomposition of checkerboard type. The raw data for a set of the
Trotter numbers $n$ were extrapolated into the $n\to\infty$ limit with a
parabolic fitting formula. We have calculated the chains
of length $L=4$, $8$, $16$, and $32$ under the periodic boundary
condition in the subspace with fixed total magnetization
$S^z=L/2$. Though the $L=32$ data are already close to the
corresponding thermodynamic limit values, we have carried out the
$L\to\infty$ extrapolation.
We have checked that almost the same
results are obtained at two temperatures, $T=0.04$ and $0.02$,
and thus we assume that $T=0.02$ is low enough to describe the
ground state properties successfully. Typically, about $10^6$ MC
steps were needed to reach the equilibrium. The ground state energy per
unit cell was estimated to be $E_{\text{QMC}}=-1.455\pm 0.001$ in the
thermodynamic limit, which agrees very well with the MP
variational result. The agreement is also good for the average
spin values, the MP calculation gives $\langle
S^z_i\rangle=0.779$ and $\langle \tau^z_i\rangle=-0.279$, to be
compared with the QMC results $0.793\pm 0.002$ and
$-0.293\pm0.002$, respectively.
In Fig.\ \ref{fig:corr} we show the results for various spin-spin
correlation functions obtained in the MP approach, together with
the QMC data. The discrepancy never exceeds 5\%, which is much
better than the corresponding SW result.
\cite{BreMikYam96}


{\em Summary.\/}
To conclude, we propose a new version of the matrix product (MP) states
approach to the description of quantum spin chains, which allows
one to construct MP states with certain quantum numbers $J$ (the
total spin) and $M$ (its $z$-projection). Known MP
representations of VBS states for integer-spin antiferromagnetic
chains,\cite{TotsukaSuzuki95} and recently proposed MP ansatz for
$S$=$1\over2$ ladder\cite{Brehmer+96} correspond to the particular case
$J=0$, $M=0$ of our ansatz. The method can be useful for the
variational description of quantum ferrimagnetic chains whose
ground state has nonzero $J$, and of higher-$S$ spin ladders.
We apply our approach to
the isotropic ferrimagnetic Heisenberg chain with
alternating spins 1 and $1\over2$; on the other hand, we study
this system numerically using the quantum Monte Carlo (QMC)
technique. For both the ground state energy and the correlation
functions we obtain very good agreement between the variational
results and the QMC data.
We think that the fundamental reason for the success of the MP
approach to the ferrimagnetic chain is that in this model the
correlations of fluctuations above the long-range order are very
short-ranged. MP ansatzes generally yield short range
correlations which is just appropriate here.

{\em Acknowledgements.\/}
It is our pleasure to thank S.~Brehmer, H.-U.~Everts, and
U.~Neugebauer for useful discussions. A.K. and S.Y. gratefully
acknowledge the hospitality of Hannover Institute for Theoretical
Physics during their stay there. This work was supported by the
German Federal Ministry for Research and Technology (BMBF) under
the contract 03MI4HAN8; one of us (A.K.) acknowledges the support
by Deutsche Forschungsgemeinschaft.

\newpage

\end{document}